# Iterated risk measures for risk-sensitive Markov decision processes with discounted cost

**Takayuki Osogami**
IBM Research - Tokyo
1623-14 Shimotsuruma, Yamato-shi
Kanagawa 242-8502, Japan

## Abstract

We demonstrate a limitation of discounted expected utility, a standard approach for representing the preference to risk when future cost is discounted. Specifically, we provide an example of the preference of a decision maker that appears to be rational but cannot be represented with any discounted expected utility. A straightforward modification to discounted expected utility leads to inconsistent decision making over time. We will show that an iterated risk measure can represent the preference that cannot be represented by any discounted expected utility and that the decisions based on the iterated risk measure are consistent over time.

## 1 Introduction

Decision making under uncertainties strongly depends on how the risk is taken into account. A popular approach is to formulate a Markov decision process (MDP) and select actions so that expected return is maximized [10, 11]. However, such risk-neutral actions can lead to huge loss with a small probability, which is not always preferable. A utility-function can be used to represent the preference to riskiness by enforcing that an undesirable outcome has a low value of utility [15]. When future reward (or cost) is discounted, the standard objective of the MDP in the literature is Discounted Expected Utility (DEU) [12], $\sum_{n=0}^{N} \lambda^n \mathsf{E}[u(R_n)]$, where $\lambda$ is the discount factor, $u$ is a utility function, $R_n$ is the reward obtained at period $n$, and $[0, N]$ denotes the planning horizon.

The most popular utility-function is of exponential form [2, 5, 6, 9, 13]. The popularity of the exponential utility-function is due to its versatility in representing the preference to riskiness and due to its computational advantage. By maximizing the expected value of an exponential utility-function of the return, a decision maker can select desirable sensitivity to risk by setting a particular value for the parameter of the exponential utility-function. Also, dynamic programming can be used to efficiently find an optimal policy with respect to the expected exponential-utility [5].

The first contribution of this paper is to demonstrate a limitation of DEU in representing the preference of a decision maker who appears to be rational. We show an example of a setting where a choice appears to be more attractive than another, but there is no utility function such that the former choice has a higher value of DEU than the latter. This limitation stems from the fact that DEU applies the utility function to the reward obtained at each period (i.e., $u(C_R)$ for each $n$) and not to the discounted cumulative reward (i.e., $u(\sum_{n=1}^{N} \lambda^n R_n)$). In other words, the DEU takes into account the risk of the reward obtained at each period, but not the risk of the discounted cumulative reward. The study of this particular limitation has not been reported in the literature.

Unfortunately, selecting actions based on $\mathsf{E}[u(\sum_{n=1}^{N} \lambda^n R_n)]$, an Expected Utility of the Discounted cumulative reward (EUD), can be problematic. We will see that decisions today can be inconsistent with decisions in future, when the decisions are made so that an EUD is maximized (particularly when the utility function is exponential). This time-inconsistency has also be observed implicitly in the prior work, which reports that the optimal policy of an MDP can change over time when its objective is to maximize, at each moment, an EUD with an exponential utility function [6, 16]. We explicitly construct a simple and clear example that illuminates time-inconsistency of EUD and discuss its negative implications.

Our second contribution is to demonstrate that the limitations of DEU and EUD can be overcome with an iterated risk measure (IRM) proposed in [1, 4]. Specifi-cally, we will show that an IRM can represent the preference of a decision maker that cannot be represented with any DEU. Also, we can guarantee that the decisions made based on the IRM are time-consistent. In addition, an optimal policy for an MDP with respect to an IRM can be found efficiently with dynamic programming, if the IRM has the properties that we refer to as strong monotonicity, translation-invariance, and positive homogeneity. We will also discuss the cases where some of the properties are not required to be satisfied for dynamic programming to apply.

The rest of the paper is organized as follows. In Section 2, we show the limitations of DEU and EUD. In Section 3, we will show that these limitations can be overcome with IRMs, where the definition of IRM is stated only informally. In Section 4, we will state a formal definition of the IRM and discuss when dynamic programming can be used to find an optimal policy with respect to an IRM. In Section 5, we will show that an IRM has advantage over expected exponential utility even when future reward is not discounted.

## 2 Limitations of standard approaches

In the following, we consider disutility of cost, instead of utility of reward, so that a decision-maker following DEU chooses the policy that minimizes $\mathsf{DEU}_{\bar{u},\lambda}(\{C_n\}_0^N) \equiv \sum_{n=0}^N \lambda^n \, \mathsf{E}\left[\bar{u}(C_n)\right]$, where $\lambda$ is a discount rate, $\bar{u}(\cdot) \equiv -u(\cdot)$ is a disutility function, and $C_n \equiv -R_n$ is the cost incurred at time $n$. We limit our discussion to a finite $N$. We choose to minimize disutility rather than maximizing utility, because we will also study risk measures, which are minimized by convention. Throughout, we assume that $\bar{u}(c)$ is increasing with $c$ (i.e., the greater the cost, the greater the disutility). Without loss of generality, let $\bar{u}(0) = 0$.

One might suspect that preference of a decision maker might be better represented with $\mathsf{EUD}_{\bar{u},\lambda}(\{C_n\}_0^N) \equiv \mathsf{E}[u(\sum_{n=0}^N \lambda^n C_n)]$ or more generally with a Risk Measure of the Discounted cumulative cost (RMD): $\mathsf{RMD}_{\mathsf{RM},\lambda}(\{C_n\}_0^N) \equiv \mathsf{RM}[\sum_{n=0}^N \lambda^n C_n]$, where $\mathsf{RM}$ denotes a risk measure, an arbitrary function that maps a random variable to a real number. Notice that EUD and RMD can represent the preference of a decision maker who is sensitive to the risk of cumulative cost rather than to the risk of the cost incurred at each period. Notice that DEU is equivalent to EUD if and only if their disutility function $\bar{u}$ is identity.

In Section 2.1, we illustrate the limitation of DEU in representing the preference of a risk-sensitive decision maker. In Section 2.2, we illustrate that EUD can lead to inconsistent decisions over time. Throughout this section, future cost is in general discounted.

|   | Day 0 | Day 1 | $\cdots$ | Day 19 | Probability |
|---|---|---|---|---|---|
| A | $1K | 0 | $\cdots$ | 0 | 1.0 |
| B | $1K | $1K | $\cdots$ | $1K | 0.0475 |
|   | 0 | 0 | $\cdots$ | 0 | 0.9525 |

Table 1: Payments with A and B. Each row shows payments for 20 days and the associated probability.

### 2.1 Limitation of DEU

We construct an example of the preference that appears rational but cannot be represented with any DEU. Consider the two payment methods, A and B, summarized in Table 1. With A, we pay $1,000 today (Day 0). With B, we might pay $1,000 on each of the consecutive 20 days starting from today, but the payment is needed only with probability 0.0475. The expected amount of total payment with B is $950, which is smaller than the $1,000 with A. However, B would require a huge amount of $20,000 with nonnegligible probability. Hence, decision makers who are averse to risk would prefer A to B, which we claim to be rational.

We will show that this rational preference (i.e., A is preferred to B) cannot be represented with any DEU. Specifically, we will show that, for any $0 \leq \lambda \leq 1$, there is no disutility function $\bar{u}$ such that

$$\mathsf{DEU}_{\bar{u},\lambda}(\{A_n\}_0^{19}) \leq \mathsf{DEU}_{\bar{u},\lambda}(\{B_n\}_0^{19}) \qquad (1)$$

where $A_n$ denotes the amount of payment with A on Day $n$ for $n = 0, 1, \ldots, 19$, and we define $B_n$ analogously. Recall that $\bar{u}(0) = 0$ and $\bar{u}$ is increasing, so that $\bar{u}_1 \equiv \bar{u}(\$1,000) > 0$. Because $\mathsf{DEU}_{\bar{u},\lambda}(\{B_n\}_0^{19})$ is increasing with $\lambda$ when $\bar{u}_1 > 0$ and $\bar{u}(0) = 0$, we have

$$\mathsf{DEU}_{\bar{u},\lambda}(\{B_n\}_0^{19}) \leq \mathsf{DEU}_{\bar{u},1}(\{B_n\}_0^{19}) = 20 \times 0.0495 \, u_1$$
$$< u_1 = \mathsf{DEU}_{\bar{u},\lambda}(\{A_n\}_0^{19}),$$

which implies that no $\bar{u}$ and $\lambda$ satisfy (1). The above argument can be summarized as follows:

**Proposition 1** *Let $A_0 = c > 0$ and $A_n = 0, \forall n \in [1, N]$. Let $\Pr(B_n = c, \forall n \in [0, N]) = (1 - \varepsilon)/(N + 1)$ and $\Pr(B_n = 0, \forall n \in [0, N]) = 1 - (1 - \varepsilon)/(N + 1)$ for $0 < \varepsilon < 1$. Then, for any $0 \leq \lambda \leq 1$, we have $\mathsf{DEU}_{\bar{u},\lambda}(\{A_n\}_0^N) > \mathsf{DEU}_{\bar{u},\lambda}(\{B_n\}_0^N)$ for any increasing (disutility) function $\bar{u}$.*

It is evident that the limitation of DEU comes from the fact that it ignores the disutility of discounted cumulative cost, taking into account only the disutility of immediate cost. We will next study EUD that can take into account the disutility of $\sum_{n=0}^N \lambda^n C_n$, the discounted cumulative cost. We will not consider discounted risk measure, $\sum_{n=0}^N \lambda^n \, \mathsf{RM}[C_n]$, which might solve the issues of DEU for particular settings but in general suffers from the drawback that it is only sensitive to the risk of immediate cost.

|   | Amount | Year | Probability |
|---|---|---|---|
| X | $1K | 1 | 0.3 |
| Y | $2K | 2 | 0.1 |

Table 2: Amount of payment, year to payment, and the probability of payment.

| Traffic condition, $\Psi$ | Normal | Busy |
|---|---|---|
| Probability | 0.9 | 0.1 |
| $T_P$ | 10 | $U[20, 80]$ |
| $T_Q$ | $U[0, 20]$ | 50 |

Table 3: Travel time along path P and path Q, where $U[a,b]$ denotes a uniform distribution with support $[a,b]$ for $a < b$.

## 2.2 Limitations of EUD

It might appear that we could use any disutility function in EUD (and any risk measure in RMD), but this turns out to be largely false for rational decision-making. In this section, we will show that the optimal policy can change simply because the time has passed if the decision is made based on an EUD. In particular, we consider the EUD whose disutility function is of an exponential form, $\mathsf{expu}^{(\gamma)}(C) \equiv \exp(\gamma C)$ for $\gamma > 0$, which is the popular and standard disutility function in the literature. The larger the $\gamma$ is, the more strongly the decision maker is averse to the risk.

Minimizing $\mathsf{E}[\mathsf{expu}^{(\gamma)}(C)]$ for $\gamma > 0$ is essentially equivalent to minimizing an entropic risk measure [3], $\mathsf{ERM}^{(\gamma)}[C] \equiv \frac{1}{\gamma} \ln \mathsf{E}[\exp(\gamma C)]$. However, ERM has additional desirable properties. For example, ERM can represent risk-averse, risk-neutral, and risk-seeking preferences in a coherent way. By minimizing $\mathsf{ERM}^{(\gamma)}[C]$ with $\gamma < 0$, the decision maker can seek risk more aggressively than by minimizing $\mathsf{E}[C]$. Although minimizing $\mathsf{ERM}^{(\gamma)}[C]$ for $\gamma < 0$ is essentially equivalent to *maximizing* $\mathsf{E}[\mathsf{expu}^{(\gamma)}(C)]$, the disutility is meant to be minimized (hence, $\mathsf{E}[\mathsf{expu}^{(\gamma)}(C)]$ is defined only for $\gamma > 0$). Furthermore, $\mathsf{ERM}^{(\gamma)}[C] \to \mathsf{E}[C]$, risk-neutral, as $\gamma \to 0$. In Section 3, we will also see that ERM can be calculated recursively. In the following, we discuss ERM, but the results also apply to EUD with exponential disutility.

We now consider the two methods of payment, X and Y (see Table 2). With X, payment is needed with probability 0.3, and we pay $1,000 a year from today if the payment is needed. With Y, payment of $2,000 is needed two years from today with probability 0.1. Let the discount rate be $\lambda = 0.92$ a year. Let $X$ denote the random variable representing the amount of payment with X. We define $Y$ analogously. A key assumption here is that no information about $X$ and $Y$ is revealed for a year. For example, immediately before a year from today, $X$ is still random ($X = \$1K$ with probability 0.3 and $X = 0$ otherwise); $X$ becomes deterministic exactly a year from today.

Consider a risk-averse decision maker, who chooses actions based on $\mathsf{ERM}^{(\gamma_0)}$ with $\gamma_0 = 0.001$. Today, he would choose Y, because $\mathsf{ERM}^{(\gamma_0)}[0.92^2 \, \gamma_0 \, Y] \approx 367$ is smaller than $\mathsf{ERM}^{(\gamma_0)}[0.92 \, \gamma_0 \, X] \approx 373$. However, if he evaluates the risk again immediately before a year from today, then he would prefer X to Y, because $\mathsf{ERM}^{(\gamma_0)}[\gamma_0 \, X] \approx 415$ is smaller than $\mathsf{ERM}^{(\gamma_0)}[0.92 \gamma_0 \, Y] \approx 425$. Note that his preference has changed simply because the time has passed. For this reason, we say that ERM is not time-consistent when future cost is discounted.

We remark that DEU is time-consistent even when future cost is discounted. For example, a risk-neutral decision maker, who makes a decision today, would choose Y, because $\mathsf{E}[0.92^2 \, Y] \approx 169$ is smaller than $\mathsf{E}[0.92 \, X] \approx 276$. A year later, she still prefers Y to X, because $\mathsf{E}[0.92 \, Y] \approx 184$ is smaller than $\mathsf{E}[X] = 300$.

## 3 IRM with discounted cost

When future reward is not discounted, ERM is known to be time-consistent [3] and also is attractive for computational purposes. A key property of ERM that we can exploit for efficient optimization is its recursiveness. We discuss the recursiveness of ERM in Section 3.1 to motivate IRM, which will be defined informally in Section 3.2 (a formal definition is postponed to Section 4). In Section 3.3, we will show that an IRM can represent the preference of a decision maker that cannot be represented with any DEU.

### 3.1 Recursiveness of entropic risk measure

Consider a two-period setting as follows. At time 0 (the day before departure), a traveler knows that $T_P$, the travel time along path P, has the distribution specified in Table 3 (we will use $T_Q$ only in Section 5). Specifically, the traffic condition, $\Psi$, is normal with probability 0.9 and busy with probability 0.1. When $\Psi$ is normal, we have $T_P = 10$. When $\Psi$ is busy, $T_P$ has a uniform distribution with support $[20, 80]$. At time 1 (immediately before departure), the traveler finds out whether $\Psi$ is normal or busy.

We can calculate $\mathsf{E}[T_P]$ in two ways. First, $T_P$ has a mass probability of 0.9 at $T_P = 10$ and a proba-bility density $f_P(x) = 1/600$ for $20 \leq x \leq 80$, so that $\mathsf{E}[T_A] = 10 \times 0.9 + \int_{20}^{80} x/600 \, dx = 14$. Second, the conditional expected value at time 1 is $\mathsf{E}[T_P|\Psi = \text{normal}] = 10$ and $\mathsf{E}[T_P|\Psi = \text{busy}] = 50$; because $\Psi$ is normal with probability 0.9 and busy with probability 0.1, we have $\mathsf{E}[T_P] = 10 \times 0.9 + 50 \times 0.1 = 14$. Formally, $\mathsf{E}[\mathsf{E}[T_P|\Psi]] = \mathsf{E}[T_P]$.

The ERM has also this recursive property. Formally, $\mathsf{ERM}^{(\gamma)}[\mathsf{ERM}^{(\gamma)}[T_P|\Psi]] = \mathsf{ERM}^{(\gamma)}[T_P]$ for any $\gamma$. For example, when $\gamma = 1.0$, we have

$$\ln(0.9\, e^{10} + 0.1\, e^{\ln \int_{20}^{80} \frac{e^x}{60} dx}) = \ln(0.9\, e^{10} + \int_{20}^{80} \frac{e^x}{600}\, dx),$$

where the left-hand side calculates $\mathsf{ERM}^{(1.0)}[T_P]$ recursively. If we evaluate $\mathsf{ERM}^{(1.0)}[T_P|\Psi]$ at time 0, we do not know whether $\Psi$ is normal or busy, so that $\mathsf{ERM}^{(1.0)}[T_P|\Psi]$ is a random variable. At time 1, however, $\Psi$ becomes known, and $\mathsf{ERM}^{(1.0)}[T_P|\Psi]$ becomes deterministic. Now, notice that a risk measure, including ERM, is a function that maps a random variable to a real number that represents the riskiness of the random variable. Then recursive calculation of $\mathsf{ERM}^{(1.0)}[\mathsf{ERM}^{(1.0)}[T_P|\Psi]]$ can be interpreted as follows. We will evaluate the riskiness of $T_P$ with $\mathsf{ERM}^{(1.0)}[T_P|\Psi]$ at time 1, and the value of $\mathsf{ERM}^{(1.0)}[T_P|\Psi]$ is random at time 0. So, at time 0, we evaluate the riskiness of the $\mathsf{ERM}^{(1.0)}[T_P|\Psi]$ with $\mathsf{ERM}^{(1.0)}[\mathsf{ERM}^{(1.0)}[T_A|C]]$.

### 3.2 Informal definition of IRM

The idea of recursive calculation of ERM motivates us to study IRM. For example, suppose that we will evaluate the conditional tail expectation (CTE) of $T_P$ at time 1. CTE is also known as conditional value at risk and formally defined as follows:

$$\mathsf{CTE}^{(\alpha)}[Y] \equiv \frac{(1-\beta)\, \mathsf{E}[Y \mid Y > V_\alpha] + (\beta - \alpha)\, V_\alpha}{1 - \alpha}, \quad (2)$$

where $V_\alpha \equiv \min\{y \mid \Pr(Y \leq y) \geq \alpha\}$ is a value at risk of a random variable $Y$, and $\beta \equiv \Pr(Y \leq V_\alpha)$. When $Y$ has a continuous distribution, we simply have $\mathsf{CTE}^{(\alpha)}[Y] = \mathsf{E}[Y \mid Y > V_\alpha]$. The 50%-CTE of $T_P$ to be evaluated at time 1 will be $\mathsf{CTE}^{(50\%)}[T_P \mid \Psi = \text{normal}] = 10$ with probability 0.9 and $\mathsf{CTE}^{(50\%)}[T_P \mid \Psi = \text{busy}] = 65$ with probability 0.1. Because $\mathsf{CTE}^{(50\%)}[T_P|\Psi]$ is a random variable, we can evaluate its 50%-CTE, which turns out to be $\mathsf{CTE}^{(50\%)}[\mathsf{CTE}^{(50\%)}[T_P|\Psi]] = 21$.

Note that $\mathsf{CTE}^{(50\%)}[\mathsf{CTE}^{(50\%)}[\cdot]]$ is a risk measure but different from $\mathsf{CTE}^{(50\%)}[\cdot]$, because $\mathsf{CTE}^{(50\%)}[T_P] = 18$. In fact, $\mathsf{CTE}^{(50\%)}[\mathsf{CTE}^{(50\%)}[\cdot]]$ is called an iterated conditional tail expectation (ICTE) in [1, 4]. Notice that an IRM, $\rho[\cdot] \equiv \dot{\rho}_0[\dot{\rho}_1[\cdot]]$, can be constructed from arbitrary risk measures, $\dot{\rho}_0$ and $\dot{\rho}_1$. Also, an IRM can be extended to more than two periods (e.g., $\rho \equiv \dot{\rho}_0 \dot{\rho}_1 \dot{\rho}_2$).

### 3.3 Overcoming the limitations with IRM

We will see, in Section 4, that ICTE satisfies the conditions that suffice to guarantee that dynamic programming can be used to find an optimal policy with respect to an ICTE when future reward is discounted. The optimal policy found with dynamic programming is guaranteed to be optimal over time, and ICTE is time-consistent in this sense.

In the rest of this section, we will see that an IRM can represent the preference that cannot be represented with any DEU. Specifically, recall the example constructed in Section 2.1, where we have seen that no DEU can represent the preference of the decision maker who would choose payment method A over B. We will show that this preference can be represented with ICTE. We will use $\mathsf{ICTE}_n^{(\alpha)}$ to denote the ICTE evaluated on Day $n$, where the ICTE is defined recursively with CTE having parameter $\alpha$ (see (2)).

First, observe that $\mathsf{ICTE}_0^{(\alpha)}[\sum_{n=0}^N \lambda^n A_n] = 1,000$ for any $\alpha$, because $\sum_{n=0}^N \lambda^n A_n = 1,000$ surely. A key property of ICTE is that a deterministic constant is mapped to the constant itself (this can be shown more formally, using ICTE's translation-invariance and positive-homogeneity defined in Section 4).

The amount of payment with B is probabilistic, and its ICTE can be calculated as follows. We define a state space, $\mathbf{S} \equiv \{0, 1\}$, so that, immediately after the beginning of Day 0, we transition to state $S = 0$ with probability 0.9525 and to $S = 1$ otherwise. The state does not change for 20 days. If $S = 0$, the payment is not needed. If $S = 1$, the payment is needed as specified in the third row of Table 1. We can calculate the ICTE for each state and for each day backwards. In particular, $\mathsf{ICTE}_0^{(\alpha)}$ is calculated from $\mathsf{ICTE}_1^{(\alpha)}$:

$$\mathsf{ICTE}_0^{(\alpha)}[\sum_{n=0}^N \lambda^n B_n] = \mathsf{CTE}^{(\alpha)} \left[ \mathsf{ICTE}_1^{(\alpha)}[\sum_{n=0}^N \lambda^n B_n \mid S] \right].$$

Because the amount of payment is deterministic given $S$, we have $\mathsf{ICTE}_1^{(\alpha)}\left[\sum_{n=0}^N \lambda^n B_n \mid S = 0\right] = 0$, and

$$\mathsf{ICTE}_1^{(\alpha)}[\sum_{n=0}^N \lambda^n B_n \mid S = 1] = \frac{1000\,(1 - \lambda^{20})}{1 - \lambda},$$

where we assume $\lambda < 1$ for simplicity. Because $S = 0$ with probability 0.9525 and $S = 1$ otherwise, we have

$$\mathsf{ICTE}_0^{(\alpha)}[\sum_{n=0}^N \lambda^n B_n] = \begin{cases} \frac{1000\,(1-\lambda^{20})}{1-\lambda} & \text{if } \alpha \geq 0.9525 \\ \frac{0.0475}{1-\alpha} \frac{1000\,(1-\lambda^{20})}{1-\lambda} & \text{otherwise.} \end{cases}$$

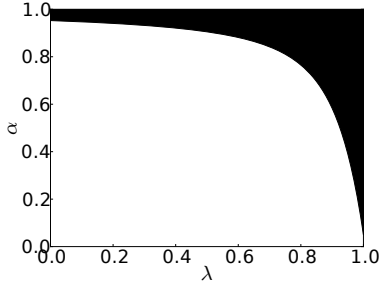

Figure 1: The painted area denotes the $(\lambda, \alpha)$ with which (3) holds. Specifically, $\alpha > 1 - 0.0475 \frac{1-\lambda^{20}}{1-\lambda}$.

Hence, we have

$$\mathsf{ICTE}_0^{(\alpha)}[\sum_{n=0}^N \lambda^n A_n] \;<\; \mathsf{ICTE}_0^{(\alpha)}[\sum_{n=0}^N \lambda^n B_n] \quad (3)$$

for sufficiently high $\alpha$ and $\lambda$. Figure 1 illustrates the region of $(\lambda, \alpha)$ where (3) holds.

We have seen that ICTE can represent the preference that cannot be represented with any DEU. Notice that we must interpret that the cost having a small ICTE is more preferable than the one with a large ICTE, which is in contrast to DEU. Otherwise, one would prefer a cost to another when the former cost stochastically dominates the latter. The above argument can be summarized as:

**Proposition 2** *Let $A_n, B_n, c, \varepsilon, N$ be as defined in Proposition 1. Let $\mathsf{ICTE}_n^{(\alpha)}$ denotes the ICTE evaluated at the beginning of the $n$-th day. Then we have $\mathsf{ICTE}_0^{(\alpha)}[\sum_{n=0}^N \lambda^n A_n] < \mathsf{ICTE}_0^{(\alpha)}[\sum_{n=0}^N \lambda^n B_n]$ if one of the following two conditions is satisfied:*

i) $\alpha > 1 - \dfrac{(1-\varepsilon)(1-\lambda^N)}{N(1-\lambda)}$ *and* $0 \le \lambda < 1$;

ii) $\alpha > \varepsilon$ *and* $\lambda = 1$.

Proposition 2 should be compared against Proposition 1. The definition of ICTE will be made formal in Section 4.

## 4 Dynamic programming with IRM

In this section, we discuss the conditions that an IRM should satisfy so that dynamic programming can be used to find an optimal policy for an MDP with respect to the IRM of the discounted cumulative cost. The optimal policy found by dynamic programming is optimal for each $n \in [0, N)$ because of the way the optimal policy is constructed. That is, the IRM satisfying the conditions to be specified in this section is time-consistent.

### 4.1 Definitions

Roughly speaking, an IRM, $\rho$, is defined recursively as $\rho_n[\cdot] = \dot\rho_n[\rho_{n+1}[\cdot]]$, where $\dot\rho_n[\cdot]$ is a risk measure (RM; more precisely, conditional RM to be defined in the following) for $n \in [0, N)$. It is important to understand $\rho_n$ and $\dot\rho_n$ in a dynamic setting. For example, $\rho_n[Y]$ represents the riskiness of $Y$ that is evaluated at time $n$, where $Y$ is in general random at time $n$. Before time $n$, $\rho_n[Y]$ is random, because the value of $\rho_n[Y]$ depends on the state $S_n$ at time $n$ and $S_n$ is random before time $n$. As time passes, we obtain more information about $\rho_n[Y]$, and $\rho_n[Y]$ becomes deterministic at time $n$. In this sense, $\rho_n[Y]$ is called $\mathcal{F}_n$-measurable, which can be understood more precisely with measure theory. Formally,

**Definition 1** *Consider a filtered probability space, $(\Omega, \mathcal{F}, P)$, such that $\mathcal{F}_0 \subseteq \mathcal{F}_1 \subseteq \ldots \subseteq \mathcal{F}_N = \mathcal{F}$, where $N \in [1, \infty)$. Let $Y$ be an $\mathcal{F}$-measurable random variable. We say that $\rho$ is an IRM if $\rho_N[Y] = Y$ and $\rho_n[Y] = \dot\rho_n[\rho_{n+1}[Y]]$, where $\dot\rho_n$ is a conditional RM that maps an $\mathcal{F}_{n+1}$-measurable random variable to an $\mathcal{F}_n$-measurable random variable, for $n \in [0, N)$.*

Notice that $\dot\rho_n$ is a conditional RM instead of a (classical) RM, which maps a random variable to a real number. For example, a conditional expectation, $\mathsf{E}[\cdot \mid S_\ell]$, is a conditional RM, where $S_\ell$ denotes the state at time $\ell$. Observe that $\mathsf{E}[X \mid S_\ell]$ is random before time $\ell$ but is deterministic at and after time $\ell$. When $\dot\rho_\ell[\cdot] = \mathsf{E}[\cdot \mid S_\ell]$ for $\ell \in [0, N)$, the corresponding IRM is also a conditional expectation, $\rho_\ell[\cdot] = \mathsf{E}[\cdot \mid S_\ell]$, for $\ell \in [0, N)$ by the recursiveness of expectation. Likewise, if $\dot\rho_\ell$ is a conditional entropic-risk-measure, $\mathsf{ERM}_\gamma[\cdot \mid S_\ell]$, for $\ell \in [0, N)$, then we have $\rho_\ell[\cdot] = \mathsf{ERM}_\gamma[\cdot \mid S_\ell]$ for $\ell \in [0, N)$. If $\dot\rho_\ell$ is a conditional CTE, $\mathsf{CTE}_\gamma[\cdot \mid S_\ell]$, $\ell \in [0, N)$, then the corresponding $\rho$ is an ICTE. Notice that IRM $\rho_n$ evaluated at a particular time $n$ is a conditional RM for each $n \in [0, N)$.

In the following, we define a class of IRMs that allow dynamic programming to find an optimal policy with respect to the IRM. Specifically, we will define strong monotonicity, translation-invariance, and positive homogeneity of the IRM. These definitions can also be found for example in [14].

If a cost $X$ is surely larger than a cost $Y$, then the value of a conditional RM of $X$ should be larger than that of $Y$. A conditional RM having such a property is called monotonic. We study an IRM defined recursively with monotonic conditional-RMs. Formally,

**Definition 2** *Consider a conditional RM, $\dot\rho$, that maps a $\mathcal{G}_1$-measurable random variable to a $\mathcal{G}_0$-measurable random variable. We say that $\dot\rho$ is monotonic if $\Pr(X \ge Y) = 1 \Rightarrow \Pr(\rho[X] \ge \rho[Y]) = 1$ for* any $\mathcal{G}_1$-measurable $X$ and $Y$. An IRM, $\rho$, as defined in Definition 1, is called strongly monotonic if $\dot{\rho}_\ell$ is monotonic for each $\ell \in [0, N)$.

For simplicity, we say that a RM is monotonic, when the corresponding conditional RM is monotonic. For example, expectation, ERM, and CTE are monotonic. Hence, ICTE is strongly monotonic. We can also compose a complex monotonic-RM from simple ones. For example, the following composite RM is monotonic:

$$\rho[\cdot] \equiv (1-\beta)\mathsf{E}[\cdot] + \beta\,\mathsf{CTE}_\alpha[\cdot] \quad (4)$$

for $0 < \beta < 1$, because

**Proposition 3** *Let $f(x_1, \ldots, x_n)$ be a non-decreasing function (i.e., $\partial f/\partial x_i \geq 0$ for $i \in [1, n]$). If conditional RM $\dot{\rho}^{(i)}$ is monotonic for each $i \in [1, n]$, then conditional RM $\dot{\rho}[\cdot] \equiv f(\dot{\rho}^{(1)}[\cdot], \ldots, \dot{\rho}^{(n)}[\cdot])$ is also monotonic.*

However, variance and other RMs that quantify the deviations from the mean, either one-sided or two-sided, are usually not monotonic. We will also use the following definitions:

**Definition 3** *Consider the $\dot{\rho}$ as defined in Definition 2. We say that $\dot{\rho}$ is translation-invariant if $\dot{\rho}[X + b] = \dot{\rho}[X] + b$ for any $\mathcal{G}_1$-measurable $X$ and $\mathcal{G}_0$-measurable $b$. We say that $\dot{\rho}$ is positive homogeneite if $\dot{\rho}[a\,X] = a\,\dot{\rho}[X]$ for any $\mathcal{G}_1$-measurable $X$ and $\mathcal{G}_0$-measurable $a$.*

In particular, expectation, ERM, CTE, and the composite RM (4) are translation-invariant. Also, expectation, CTE, and the composite RM (4) are positive homogeneite, but ERM is not.

### 4.2 Dynamic programming

Consider a discrete time MDP over a finite horizon $[0, N]$, having state space $\cup_{n \in [0,N]}\mathbf{S}_n$, action space $\mathbf{A}$, transition probability $p_n(s_{n+1}|s_n, a_n)$, and cost $r_n(s_{n+1}, s_n, a_n)$ for $s_n \in \mathbf{S}_n, s_{n+1} \in \mathbf{S}_{n+1}, a_n \in \mathbf{A}$, and $n \in [0, N)$. Here, $p_n(s_{n+1}|s_n, a_n)$ denotes the probability that the state at time $n+1$ is $s_{n+1}$ given that action $a_n$ is taken from state $s_n$ at time $n$; $r_n(s_{n+1}, s_n, a_n)$ denotes the cost incurred immediately after time $n$ when action $a_n$ is taken from state $s_n$ at time $n$ and the state at $n+1$ is $s_{n+1}$. For succinctness, $r_n(s_{n+1}, s_n, a_n)$ will be denoted by $r_n$.

A policy, $\pi$, is a function that maps a pair of a time, $\ell$, and a state, $s_\ell$, into an action, $a_\ell$ (i.e., the action $a_\ell$ is taken at time $\ell$, if the state at time $\ell$ is $s_\ell$). Let $\Pi$ be the set of candidate policies.

We seek to find an optimal policy, $\pi^\star \in \Pi$, that minimizes an IRM, $\rho_n[R^{(\lambda)}|\pi]$, of the discounted cumulative cost, $R^{(\lambda)} = \sum_{\ell=0}^{N-1} \lambda^\ell\,r_\ell$, at every $n \in [0, N)$. When $\rho_\ell$ is translation-invariant and positive homogeneite, we have

$$\rho_\ell[R^{(\lambda)}] = \sum_{i=0}^{\ell-1} \lambda^i\,r_i + \lambda^\ell\,\rho_\ell[\sum_{i=\ell}^{N-1} \lambda^{i-\ell}\,r_i]. \quad (5)$$

That is, the riskiness of $R^{(\lambda)}$ evaluated at time $\ell$ is equal to the sum of the discounted cumulative cost incurred before time $\ell$ and the riskiness of the discounted cumulative cost to be incurred after time $\ell$. Notice that the action to be taken at (and after) time $\ell$ cannot affect the cost incurred before time $\ell$. Hence, the optimal action at time $\ell$ that minimizes $\rho_\ell[\sum_{i=\ell}^{N-1} \lambda^{i-\ell} r_i]$ also minimizes $\rho_n[R^{(\lambda)}]$. That is, the optimal action at time $\ell$ is independent of the cost incurred before then.

The following theorem implies that the optimal policy can be found with dynamic programming, where the optimal policy minimizes $\rho_n[\sum_{\ell=n}^{N-1} \lambda^{\ell-n} r_\ell]$ (hence, also $\rho_n[R^{(\lambda)}]$) for every pair of state and time (given the state and the time):

**Theorem 1** *Consider the MDP as defined at the beginning of Section 4.2. Let $\rho$ be an iterated RM as defined in Definition 1. We assume that $\rho$ is strongly monotonic and that $\rho_n$ is translation-invariant and positive homogeneite for each $n \in [0, N)$. Let $V_n^\lambda(s_n, \pi_n) \equiv \rho_n[\sum_{\ell=n}^{N-1} \lambda^{\ell-n} r_\ell \mid S_n = s_n, \pi_n]$, for $n \in [0, N]$, be the value of $\rho_n[\sum_{\ell=n}^{N-1} \lambda^{\ell-n} r_\ell]$ given that we use policy $\pi_n \in \Pi$ to choose actions, starting from state $s_n \in \mathbf{S}_n$. Then, for every $s_n \in \mathbf{S}_n$ and for every $n \in [0, N)$, the solution to the following optimality equations minimizes $V_n^\lambda(s_n, \pi_n)$, where we define $V_n^\lambda(s_n) \equiv \min_{\pi_n \in \Pi} V_n^\lambda(s_n, \pi_n)$:*

$$V_N^\lambda(s_N) = 0,$$

$$V_n^\lambda(s_n) = \min_{a \in \mathbf{A}} \rho_n \left[ \mathcal{D}_{s' \in \mathbf{S}_{n+1}} \left\{ \begin{matrix} r_n(s', s_n, a) + \lambda\,V_{n+1}^\lambda(s') \\ p_n(s'|s_n, a) \end{matrix} \right\} \right]$$

*for $s_\ell \in \mathbf{S}_\ell$ such that $\ell \in [0, N]$ and for $0 \leq n < N$, where $\mathcal{D}_{s \in \mathbf{S}} \left\{ \begin{matrix} x(s) \\ p(s) \end{matrix} \right\}$ denotes a discrete random variable taking value $x(s)$ with probability $p(s)$ for $s$ in a set $\mathbf{S}$.*

A proof of the theorem is provided in the associated technical report [7], where the key property that we exploit is (5).

Notice that CTE is monotonic, so that ICTE is strongly monotonic. Also, it can be shown that $\mathsf{ICTE}_n$ is translation-invariant for each $n \in [0, N)$, because an IRM defined recursively with translation-invariant $\bar{\rho}$ is translation-invariant (see Lemma 3 in the associated technical report [7]) and CTE is translation-invariant. Likewise, $\mathsf{ICTE}_n$ is positive homogeneite, because CTE is positive homogeneite (see Lemma 4 in [7]).

Notice that, when future cost is not discounted ($\lambda = 1$), positive homogeneity is not needed for (5) to hold. In fact, the proof of Theorem 1 implies that the theorem holds without positive homogeneity when $\lambda = 1$. We have seen in Section 2 that ERM is not time-consistent when future cost is discounted, even though it is time-consistent without discounting. Now, this can be explained by the fact that ERM is strongly monotonic and translation-invariant but not positive-homogeneite.

When an IRM is not translation-invariant, we cannot separate the discounted cumulative reward that has already been incurred from that to be incurred as in (5). This suggests that the optimal action depends on the discounted cumulative cost that has been incurred by the time the action is taken. Then the state of an MDP would need to include the information about the discounted cumulative cost that has already been incurred [8]. The expanded state space would make the optimization with dynamic programming less efficient. For this reason, ERM might appear to be an ideal RM when future cost is not discounted, because ERM is strongly monotonic and translation-invariant. However, we will see in Section 5 that ERM cannot represent some of the preferences that can be represented with ICTE even when future cost is not discounted.

## 5 Undiscounted future cost

In this section, we will illustrate the limitation of ERM when future cost is not discounted. Recall the path P having the travel time $T_P$ specified in Table 3. We will also consider the path Q with travel time $T_Q$, which is uniformly distributed between 0 minutes and 20 minutes when the traffic condition is normal, and is 50 minutes when it is busy. Observe that P and Q have the same expected travel time, $\mathsf{E}[T_P] = \mathsf{E}[T_Q]$.

Informal interviews with our colleagues suggest that some people prefer path P to path Q, and others prefer Q to P. There are also people whose preference depends on the purpose of the travel and other conditions. Whether a traveler takes P or Q depends on how the traveler perceives the risk associated with the travel time.

It turns out that a traveler who makes decisions based on $\mathsf{ERM}^{(\gamma)}$ would choose path Q no matter how he sets the parameter, $\gamma$. Formally, $\mathsf{ERM}^{(\gamma)}[T_P] \geq \mathsf{ERM}^{(\gamma)}[T_Q]$ for any $\gamma$, where the equality holds only when $\gamma = 0$:

**Lemma 1** *Let $U[a, b]$ denotes a random variable uniformly distributed between $a$ and $b$ for $-\infty < a \leq b <$*

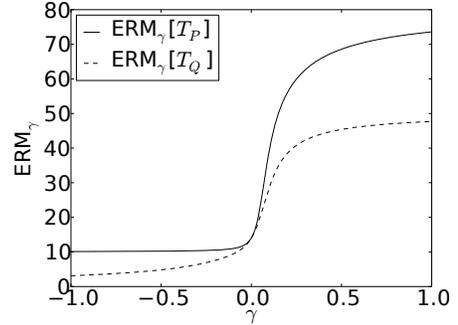

Figure 2: Entropic risk measure of $T_P$ and $T_Q$.

$\infty$. *Let $X$ and $Y$ be random variables such that:*

$$X = \begin{cases} -\frac{1}{2} & w.p. \ p \\ U[0, x] & o.w. \end{cases} \qquad Y = \begin{cases} U[-1, 0] & w.p. \ p \\ \frac{x}{2} & o.w. \end{cases}$$

*where $p \equiv x^2/(1 + x^2)$ and $0 \leq x < \infty$ Then $\mathsf{ERM}^{(\gamma)}[\alpha X + \beta] \geq \mathsf{ERM}^{(\gamma)}[\alpha Y + \beta]$ for any $\alpha > 0$, $-\infty < \beta < \infty$, and $-\infty < \gamma < \infty$, where $\mathsf{ERM}^{(0)} \equiv \mathsf{E}$.*

A proof of the lemma is provided in the associated technical report [7] (see Figure 2(c) for a numerical support). A traveler could make decisions so that he chooses a path that has a larger value of ERM. Then he would choose path P. However, such a traveler would choose a path that surely takes longer than another, because $\mathsf{ERM}^{(\gamma)}[T + c] = \mathsf{ERM}^{(\gamma)}[T] + c$ for any constant $c$ (i.e., we can make a travel time, $T$, more attractive to the traveler by adding extra travel time, $c > 0$, to the $T$). That is, maximizing ERM is by no means rational.

It turns out that ICTE can represent the preference of P to Q. Observe that $\mathsf{CTE}_1^{(50\%)}[T_Q|\Psi = \text{busy}] = 50$ and $\mathsf{CTE}_1^{(50\%)}[T_Q|\Psi = \text{normal}] = 15$, so that $\mathsf{CTE}_0^{(50\%)}[\mathsf{CTE}_1^{(50\%)}[T_Q|\Psi]] = 22$, which is greater than $\mathsf{CTE}^{(50\%)}[\mathsf{CTE}^{(50\%)}[T_P|\Psi]] = 21$. Thus, an iterated CTE can indeed represent the risk-sensitivity of the traveler who prefers path P to path Q. A CTE can also represent such risk-sensitivity, because $\mathsf{CTE}^{(80\%)}[T_p] = 30$ is smaller than $\mathsf{CTE}^{(80\%)}[T_Q] = 34 + 4/9$. However, CTE is not time-consistent and causes the issues that we have seen in Section 2.2.

## 6 Concluding remarks

We remark that the results about Theorem 1 are related to but different from Ruszczyński [14], who proposes ways to find optimally risk-averse policies for an MDP by defining a Markov risk measure and a discounted measure of risk. Specifically, Ruszczyński shows that dynamic programming can be used to find an optimal policy for an MDP over a finite horizon with respect to the Markov risk measure, where future reward is not discounted. Also, it is shown that value iteration can be used to find an optimal policy for an MDP over the infinite horizon with respect to the discounted measure of risk, where future reward is discounted. The Markov risk measure and the discounted measure of risk are defined to be particular IRMs that have specified properties, including the three properties that we require in Theorem 1. However, one of the specified properties in [14] is convexity, which we do not require in Theorem 1. We have also seen that positive homogeneity is not needed for dynamic programming to apply when future reward is not discounted.

The primary message of this paper is that DEU has limitations in representing the preference of a decision maker who is averse to or seeking risk when future reward is discounted, and such limitations can be overcome with IRM. Theorem 1 clarifies the conditions that suffice to guarantee that an optimal policy with respect to an IRM can be found with dynamic programming. These conditions in turn guarantee that the IRM is time-consistent in the sense that the optimal policy found with dynamic programming does not change over time. We have also seen that EUD (particularly, with exponential utility) is not time-consistent, and the decisions made based on EUD can be completely irrational.

The focus of this paper has been on what can represent *rational* preferences on risk and time. However, corresponding psychological studies would also be an interesting direction. There has been limited work on behavioral models that take into account both risk and time. In particular, how well can DEU and EUD represent people's preference?

## Acknowledgments

This work was supported by "Promotion program for Reducing global Environmental loaD through ICT innovation (PREDICT)" of the Ministry of Internal Affairs and Communications, Japan. The author thanks Dr. Tetsuro Morimura for fruitful discussion.